\documentstyle[12pt, subeqnarray]{article}

\begin{document}
\newcommand{\dis}{\displaystyle}
\newcommand{\beq}{\begin{equation}}
\newcommand{\eeq}{\end{equation}}
\newcommand{\noi}{\noindent}
\newcommand{\expon}{\mbox{\rm e}}

\thispagestyle{empty}
\rightline{} 
\rightline{} 

\vskip 0cm
\centerline{\large \bf Maximum temperature for an Ideal Gas
}
\centerline{\large \bf of $\hat U
(1)$ Kac-Moody Fermions\footnote{To appear in Physical Review,Dec 15, 1995.} }


\vskip 2cm
\centerline{ Belal E. Baaquie \footnote{E-mail
address: phybeb@leonis.nus.sg}}
\centerline{{\it Department of Physics, Faculty of Science, } }
\centerline{{\it National University of Singapore, Kent Ridge,} }
\centerline{{\it Singapore 0511. } }
\vskip 2in


\centerline {\em Abstract}

{ \em A lagrangian for gauge fields coupled to fermions with the Kac-Moody
group as its
 gauge group yields, for the pure fermions sector, an ideal gas of Kac-Moody
fermions.
 The canonical partition function for the $\hat U(1)$ case is shown to have a
maximum
 temperature $kT_{M} = |\lambda| /\pi$, where $\lambda $
 is the coupling of the super charge operator $G_0 $ to
 the fermions. This result is similar to the case of strings but
 unlike strings the
 result is obtained from a well-defined lagrangian.
}

\newpage

\noi {\bf Introduction}

\bigskip

 The existence of a maximum temperature is widely supposed to hold in string
 theory. In this paper we discuss another case where the same phenomenon
occurs, and
 the result is shown to hold for a model with an exact lagrangian.

 Gauge fields coupled to fermions having an arbitrary Kac-Moody group for
 its gauge symmetry has been derived in Ref 1, and has a number of
 new features. The pure
 gauge sector is nonlinear even for $\hat U (1)$ case.
 The fermion sector has a new mass-
 like coupling to the super-charge operator $G_0$ due
 to the necessity of attenuating the
 high mass states inside Feynman loop integrations (Ref 1).

 In this paper, we examine the pure fermion sector. The simplest case of
 $\hat U (1)$
 Kac-Moody fermions is studied and we derive the existence of a maximum
temperature
 for the free energy.

 Consider a $d$-dimensional Euclidean space time $M_{d}$. Let $Q_{n}$
 and $h_{n}$ be the
 generators of $\hat U (1)$ super Kac-Moody algebra with

\beq
\lbrack Q_n, Q_m \rbrack = ik \delta_{n + m}
\eeq

\beq
\lbrack h_n, h_m \rbrack = k\delta_{n + m}
\eeq

 We consider only the Ramond sector (integer modes $h_{n}).$
 Hence we have the
 Virasoro generator

\beq
L_o = \frac{1}{2k} Q^2_0 + \frac{1}{k} \sum_{n=1}^{\infty} (Q_{-n}Q_n +
n h_{-n}h_n) + \frac{1}{16}
\eeq

 and the super-charge operator

\beq
G_0 = \frac{1}{k} Q_0 h_0 + \frac{1}{k} \sum_{n=1}^{\infty} (Q_{-n}h_n +
h_{-n}Q_n)
\eeq

\noi  with

\beq
G_0^2 = L_0 - \frac{c}{24}.
\eeq

 These generators, together with $L_{n}$ \& $G_{n},$ form the $N = 1$
 super Virasoro algebra with  $c = 1 + 1/2 = 3/2$.

 \bigskip

\noi {\bf Lagrangian}

\bigskip

 The free Kac-Moody fermion lagrangian is given by [Ref 1]

\beq
L = \bar \psi(x)(\partial_\mu \gamma^\mu + \lambda G_0 + m +
\mu\gamma_0) \psi(x)
\eeq

\noi where $\lambda $ and m have dimension of mass and $\mu $ is the chemical
potential.

\noi Under gauge transformations we have [Ref 1]

$$
\psi \rightarrow \Phi\psi  , \hspace{2cm} \bar\psi \rightarrow
\bar\psi \Phi^{\dagger}
$$

\noi where

\begin{displaymath}
\Phi (x) = \exp \{i\sum^{ +\infty }_{ n=-\infty }\phi_{n}(x)Q _{n} \}
\end{displaymath}

\noi  is an element of the Kac-Moody group and $\phi_{n}(x)$
 is an infinite collection of gauge functions.

 The partition function is given by $\dis (\beta  = \frac{1}{kT})$

\begin{eqnarray}
Z & = & \mbox{\rm tr} ~ \exp\{ -\beta(H - \mu N)\} \\
& & \nonumber \\
& = & \int D\bar \psi D\psi \exp \{\int_0^\beta dt \int d^{d-1}x L \}.
\end{eqnarray}

$\psi , \bar\psi $ are arbitrary elements of an irrep of the super
Virasoro algebra. We have  commuting operators $h_{0}, ~ Q _{0}, ~
G_{0}$ and $L_{0}$.  A vacuum state is specified by $|q, h >$ which
 is annihilated by all $Q _{n}, ~ h _{n}, ~ n > 0$ and where
$Q _{0} |q, h > = q|q, h >$ and $L _{0}|q, h > =  h |q, h >$.
We choose $q = 0$, and since we are in the Ramond sector we have
$h = 1/16$; we can consequently choose a chiral representation with a
supersymmetric vacuum given  by $G_{0}|q, h > = 0$,
and this is possible since for our case $c/24 = 1/16$ [Ref 2].

 In effect, all the states of the representation space are created by $Q_{-
n}, ~ h _{- n}, ~ n > 0 $,
acting on $|q, h > = |0, 1/16 >$ and the operators $Q _{0}, ~ h _{0}$
are set to zero  everywhere.

 Hence we have the expansion for the fermion field

\beq
\psi(x) = \psi(x_0) + \psi_n(x) h_{-n} + \psi_{nm}(x)h_{-n}h_{-m} +
\psi_{nml}(x)h_{-n}h_{-m}h_{-l} + \cdots
\eeq

\noi where $\psi _{0}, \psi_{n}, \cdots $  are fermionic degrees of
freedom which carry an irrep of the  bosonic Kac-Moody algebra with
$q = 0$. From eqn (9) we see that $\psi (x)$ is similar to a
 superstring field with the $\psi_{0}, \psi_{n}$'s etc
being the excited string modes.

 The expansion given in eqn (19) is not suitable for performing the path
 integration in eqn (8). We instead choose a basis for $\psi , \bar\psi$
to diagonalize $ L$ in eqn
 (6) and hence evaluate Z exactly. Let $|n>$
be an eigenstate of $L_{0}$ with
 $L _{0}|n > = (n + \frac{ 1}{ 16}) |n >, ~ n = 0, 1, 2 ... \infty .$

 The degeneracy of each state $|n>$ is given by $d _{n}$ where [Ref 3]

\beq
\prod_{n=1}^{\infty} \frac{(1 + q^n)}{(1 - q^n)} = \sum_{n=0}^\infty
d_n q^n = \mbox{\rm tr ~ ~} q^{L_0 - \frac{c}{24}}.
\eeq

\noi The numerator on the left hand side comes from the fermionic
states at level $n$ and the  denominator from the bosonic ones.

\noi Define, for $n\neq 0,$

\begin{subeqnarray}
|n, \pm> &=& (\sqrt{n} \pm G_0) |n> \\
\mbox{\rm \hspace{-2cm} with \hspace{2cm}}  G_0 |n, \pm> & = &  \pm \sqrt{n} |n
, \pm>
\end{subeqnarray}

 We have the expansion

\beq
\psi(x) = \psi_0(x) |0> + \sum_{n=1}^\infty \sum_{\pm} \psi_n^{\pm}(x)
|n, \pm>
\eeq

\noi and similarly for $\bar\psi(x)$.

 Then, from eqns (6) and (12)

\begin{eqnarray}
 L & = & \bar\psi_{0}(x)(\partial  + m + \mu \gamma_{0})\psi_{0}(x)
\nonumber \\
& & \mbox{\hspace{2cm}} + \sum_{n=1}^\infty \sum_{\pm}
\bar\psi_{n}(x)(\partial  \pm \lambda \sqrt{n}
+ m + \mu \gamma_{0})\psi_{n}(x).
\end{eqnarray}

 We see from eqn(13) that the zero mode's $\bar\psi_{0}(x), \psi_{0}(x)$
 reproduce the ordinary
 Dirac lagrangian, and the higher modes give an infinite collection of Dirac
particles
 with masses increasing as $\mid m \pm  \lambda \sqrt{n} \mid$.

\bigskip

\noi {\bf Partition Function, Maximum Temperature}

\bigskip

 Since all the degrees of freedom are decoupled, the path integral in eqn (8),
 for each mode $\psi  _{n}^{+}(x),$ is an ideal gas of fermions with
 mass $\mid m \pm \lambda n \mid $ and can be
 evaluated exactly using standard techniques [Ref 4]. Performing the path
 integrations yields the free energy

 \begin{eqnarray}
 F &=& \ln Z/V \nonumber \\
 & = & 2^{d/2 -1}\sum_{n=0}^{\infty} d_n \sum_{\pm} \int_p \{ \ln (1 +
\expon^{\mu\beta} \expon^{-\beta\sqrt{p^2 + (m \pm \lambda \sqrt{n})^2}})
\nonumber \\
& & \mbox{\hspace{3.5cm}} + \{ \ln (1 +
\expon^{-\mu\beta} \expon^{-\beta\sqrt{p^2 +
(m \pm \lambda \sqrt{n})^2}} \}
\end{eqnarray}
\noi where  $V= \int d^{d - 1}x$,
$\dis \int_p \equiv \int \frac{d^{d-1}p}{(2\pi)^{d-1}}$,
and $d_{0} \rightarrow d_{0} / 2$.

\vspace{5mm}

  We expect $T_{M}$ to be determined by very high mass states, and hence to
leading
 order we can set $m = 0$; we will consider the case where $kT _{M}>>\mu $ and
hence to leading
 order we can also set $\mu =0$. We further consider the case of $\lambda $
such that $\beta \lambda  >> 1$ and
 this will in effect make $T _{M}$ small; we consequently study this
temperature regime
 of F to ascertain the existence of a maximum temperature.

 Doing a low temperature expansion for $F$ ,
ignoring $m$ and $\mu$, yields [Ref 4]

\begin{eqnarray}
F & = &  2^{d / 2 + 1} \sum^{ \infty }_{ n=0} d_{n}
\sum_{\pm} \int_{p} \expon^{-\beta\sqrt{p^{2} + \lambda^{2}n}} + \cdots
\nonumber \\
& = & \frac{2}{\pi^{d/2}}(\frac{\lambda}{\beta})^{\frac{d - 2}{2}}
\sum_{n=0}^{\infty} d_n n^{\frac{d-2}{2}}
K_{(d-2)/2}(\beta|\lambda|\sqrt{n}) + \cdots
\end{eqnarray}

\noi  where $K _{(d - 2) / 2}$
is the associated Bessel function and its asymptotic expansion yields
($a=$const)

\beq
F = a \sum_{n=1}^\infty d_n n^{(d - 3)/4}
\expon^{-\beta|\lambda|\sqrt{n}} .
\eeq

\noi Clearly the value for $d_{n}$ as   $n \rightarrow \infty $
will determine if $F$ is finite for all $\beta$. To determine $d _{n}$,
note, for $q \rightarrow  1$ [Ref 5, 6]

\beq
\prod_{n=1}^\infty \frac{(1 + q^n)}{(1 - q^n)} = \theta_4^{-1} (0|q)
\simeq (1 - q)^{1/2} \exp\{ - \frac{\pi^2}{4(1 - q)} \}
\eeq

\noi  where $\theta  _{4}(0\mid q)$ is the Jacobi theta function.
We perform a contour integral using eqn(10) to evaluate $d _{n}$;
there is a sharp maxima of the integrand at $q=1$, and using the
 saddle-point method with eqn (17) yields

\beq
d_n \simeq \frac{1}{n} \expon^{\pi\sqrt{n}}
\eeq

\noi  Hence

\beq
F \simeq \sum_{n=1}^\infty n^{(d-7)/4} \expon^{\pi \sqrt{n}}
\expon^{-\beta|\lambda|\sqrt{n}}.
\eeq

\noi  We see that F diverges for all $\beta \mid\lambda \mid < \pi $;
hence F is finite for all $kT < kT_{M}$, where the
 maximum temperature is $kT_{M} = \mid\lambda \mid/\pi$.

As $T \rightarrow T_M$ from below, we have from eq.(19)

\beq
F \propto \cases{\frac{1}{(T_M - T)^{d - 3}}, &  \mbox{$d > 3$} \cr
\ln(T_M - T), & \mbox{$d = 3$} \cr
\mbox{\rm finite}, & \mbox{$d < 3$} \cr}
\eeq

\bigskip

\noi {\bf Discussion}

\bigskip

 An obvious similarity of Kac-Moody fermions and strings is the exponentially
growing
 density of states, which in both cases destabilizes the system at high enough
 temperature. There are however a number of dissimilarities between the two
cases.

 Firstly a finite maximum temperature exist for Kac-Moody fermions because of
the
 competition between the square root of relativistic energy $\sqrt{ p ^{2} +
\lambda^{2}n}$ and the square
 root in entropy $\pi \sqrt{ n}.$ If for example we had used $L _{0}$ instead
of $G_0$ in the lagrangian
 $ L$, the free energy would have been convergent for all temperature.

 Secondly, Kac-Moody fermions have a well defined path integral for all
temperature and hence we can study the singularity at $T _{M}$ in
detail using {\em L }of eqn (6), whereas in the case of strings short
distance effects due to gravity are supposed to invalidate concepts
such as maximum temperature, phase transitions etc [Ref 6].
Nevertheless, studies of an ideal gas of strings have been carried out
and lead to results that differ significantly from the point particle
case [Ref 7].

 $T _{M}$ in general is a complicated function of $\lambda ,$ and we
obtained the leading behaviour of $T _{M}$ for the region $\mu /\lambda
<< 1  << \beta \lambda$. Given the exponential growth of the density of
states, one expects that the free energy F will have a
singularity at $T = T _{M} $ as in eq.(20); a detailed study of F near $T _{M}$
including the nature of the phase transition at $T _{M}$ can be
obtained from eqn (14). It can also be determined, using the lagrangian
{\em L }in eqn (10), whether the canonical and microcanonical ensembles
are equivalent for this system which has an exponentially growing
density of states.

 In conclusion, an ideal gas of Kac-Moody fermions provides an exact
model to study the physics of a system with a finite maximum
temperature.

\bigskip

\noi {\bf References}

\bigskip

 1. B. E. Baaquie, Phys. Lett. {\bf 271B }(1991) 343.

 2. C. Itzykson \& J-M Drouffe, {\em Statistical Field Theory Vol II
}(Cambridge Univ Press

 1989).

 3. S. Ketov, {\em Conformal Field Theory }(World Scientific, 1995).

 4. J I. Kaputsa, {\em Finite Temperature Field Theory }(Cambridge Univ Press,
1989).

 5. K. Chandrasekharan, {\em Elliptic Functions }(Springer Verlag, 1985).

 6. M. B. Green, J. H. Schwarz \& E. Witten {\em Superstring Theory Vol I \&
Vol II
}

 (Cambridge Univ Press, 1987).

 7. N. Deo, S. Jain \& C-I Tan, Phys. Lett. {\bf 220B }(1989) 125, Phys.Rev.
{\bf D40 }(1989) 2626.

\end{document}